\documentclass{optica-article}

\journal{opticajournal} 

\articletype{Research Article}

\begin{document}

\title{Self-focusing of high-intensity beams with grid structures}

\author{Jiaqi Wang,\authormark{1,*} Yang Xu,\authormark{2} Saumya Choudhary,\authormark{3} Omid Mozafar,\authormark{4} and Robert W. Boyd \authormark{2,3,4}}

\address{\authormark{1}CREOL, The College of Optics and Photonics at the University of Central Florida, Orlando, Florida 32816, USA\\
\authormark{2}Department of Physics and Astronomy, University of Rochester, Rochester, New York, 14627, USA\\
\authormark{3}The Institute of Optics, University of Rochester, Rochester, New York, 14627, USA\\
\authormark{4}Department of Physics, University of Ottawa, Ottawa, Ontario K1N 6N5, Canada}

\email{\authormark{*}jqwang@ucf.edu} 


\begin{abstract*} 
High-power laser beams propagating in a Kerr medium can undergo catastrophic self-focusing when their optical power exceeds a critical value set by the medium's optical properties. The resulting strong intensity localization can trigger additional nonlinear processes and cause irreversible damage. Here, we study a coherent transverse grid beam consisting of an $N\times N$ array of identical Gaussian beamlets and show that even in the absence of absorption or higher-order nonlinearities, this spatial structuring can raise the effective collapse threshold over a fixed propagation distance within a non-paraxial Kerr propagation model. The key mechanism is Kerr-mediated coupling between neighboring beamlets, which redistributes power across the array and delays the localized collapse. We find that above the collapse threshold the nonlinear evolution proceeds through an N-dependent multi-stage coalescence cascade, and that the maximum non-collapsing transmitted power is achieved at an optimal lattice spacing. We find that square grids of optimal lattice spacing can propagate $\approx$ 44\% more power than the total critical power for the grid with isolated beamlets without undergoing a collapse within one Rayleigh range. Finally, we provide an empirical scaling relation linking the optimal lattice spacing to the radius of the beamlet. Our proposed simple beam structuring strategy has implications for  mitigating self-focusing in high-power transmission and directed-energy applications.
\end{abstract*}

\section{Introduction}
Self-focusing is a common nonlinear optical phenomenon observed in many high-intensity laser systems \cite{askar1974self, shen1975self, boyd2009self}. In a Kerr medium, the refractive index depends on intensity as $n = n_0 + n_2 I$, where $n_0$ is the linear refractive index of the material, $n_2$ is the third-order nonlinear coefficient of refractive index, and $I$ is the intensity of the propagating beam \cite{boyd2020nonlinear}. For $n_2 > 0$, the induced index profile acts as a focusing lens for beams with a central intensity maximum, which can lead to a runaway increase of the on-axis intensity in the absence of competing mechanisms such as plasma defocusing. While self-focusing can be exploited in applications such as supercontinuum generation \cite{alfano1970observation} and nonlinear beam combining \cite{lushnikov2014nonlinear}, it also amplifies phase noise and modulational instability, promoting filamentation, beam breakup \cite{brewer1968small, couairon2007femtosecond, boyd2020nonlinear}, rogue-wave/caustic formation \cite{safari2017generation, PhysRevLett.129.133902, choudhary2024controlling}, and ultimately material damage \cite{soileau1989laser}. These deleterious effects impose a practical upper bound on the power that can be transmitted with a stable spatial profile over a required distance. 

Many applications such as long-range atmospheric propagation \cite{nelson2016atmospheric, rouze2021coherent}, free-space optical communication \cite{malik2012influence, zhu2021compensation}, and laser material processing \cite{bi2023beam} require the delivery of high optical power while maintaining beam quality. In the early stages of collapse, a Gaussian beam evolves toward a Townes-like profile \cite{chiao1964self, moll2003self}, shedding energy from the central region. In experiments, this “shed” energy can appear as a redistribution into broader spatial components; in numerical propagation it is often accompanied by radiation into high-transverse-wavevector components. In either case, this redistribution and the onset of small-scale filamentation can substantially degrade beam quality and reduce the usable transmitted power. Consequently, significant effort has been devoted to mitigating collapse and controlling nonlinear propagation, including dispersion management \cite{kivshar2003optical, TURITSYN2012135}, spatial mode engineering \cite{PhysRevA.77.043814, PhysRevA.100.013836}, and polarization structuring \cite{bouchard2016polarization, PhysRevLett.129.133902, Lu:22, wang2023self}. Even in the simple case of two co-propagating Gaussian beams, collapse dynamics can be tuned through transverse separation, relative phase, and power \cite{ishaaya2007self, shim2010controlled}.

A standard benchmark is the critical power for catastrophic self-focusing. For Gaussian beams, Fibich and Gaeta \cite{Fibich:00} give a quantitative expression
\begin{equation}
	\label{P0}
	P_{0} = 1.8962 \frac{\lambda^2}{4\pi n_0 n_2},
\end{equation}
where $\lambda$ is the wavelength in the free space. In the paraxial nonlinear Schrodinger equation, catastrophic collapse is associated with a singularity, which is unphysical and complicates analysis beyond the first focusing event. Importantly, previous works \cite{cmp/1103922134,MALKIN1993251} have shown that including weak non-paraxiality can arrest this singularity through optical effects alone.

Here, we investigate the role of transverse power redistribution as a practical handle to delay collapse. Specifically, we consider a coherent $N\times N$ grid of Gaussian beamlets with uniform phase and equal power per beamlet, and we model propagation in the non-paraxial regime. We show that, for a fixed propagation distance $z_{\text{max}}$, a grid with an appropriate lattice spacing can transmit a total power exceeding the fully separated-beamlet limit $N^2 P_0$ by leveraging Kerr-mediated coupling among neighboring beamlets. This coupling produces interference-driven modulation of the nonlinear index and enables lateral power exchange across the array, thereby delaying localized collapse. For square grids, we find a distance-dependent enhancement of up to 44\% at $z_{\max}=z_{\rm R}$ (15\% at $z_{\max}=4z_{\rm R}$) in the total power that can be carried by the beam without undergoing collapse, or the "non-collapse power", at an optimal lattice spacing. Finally, we extract an empirical scaling relation between the optimal spacing and the beamlet radius for a range of grid sizes and layouts.

\begin{figure}[htbp]
	\centering\includegraphics[width=8cm]{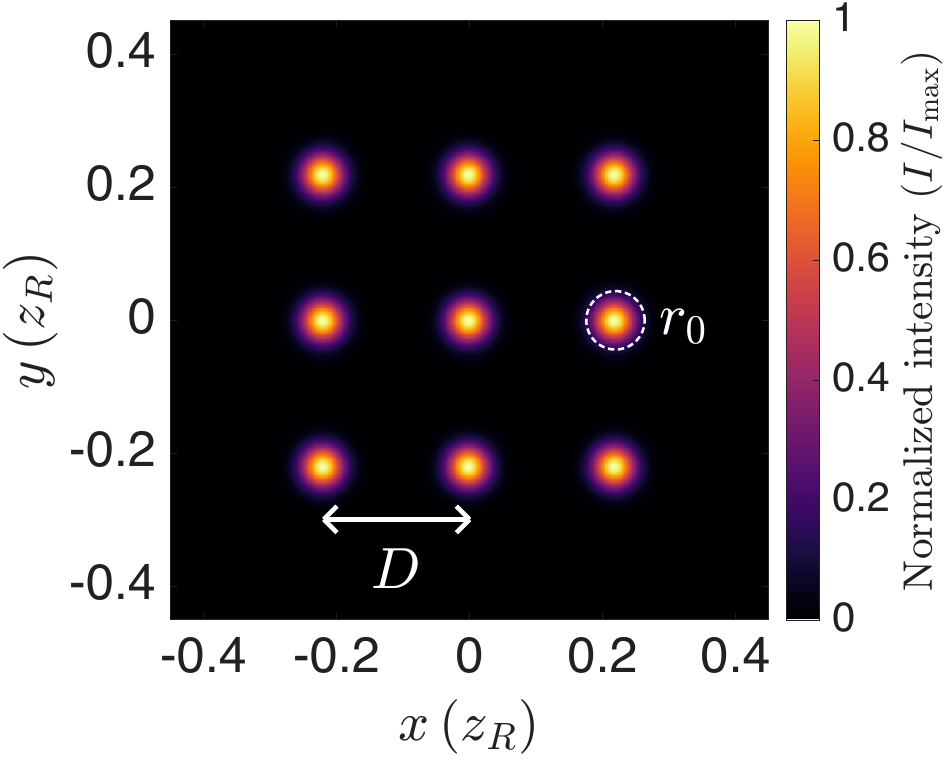}
	\caption{Intensity profile at the entrance surface of an $N \times N$ grid beam ($N = 3$). The spatial coordinates are scaled to the Rayleigh range $(z_R)$ of each Gaussian beamlet.}
	\label{f1}
\end{figure}

\section{Nonlinear non-paraxial propagation of grid structures}
Figure \ref{f1} shows the spatial intensity distribution of an example $N \times N$ grid of Gaussian beams with $N = 3$ considered in our study. For simplicity, we assume that the total power of the beam is evenly distributed across all $N^2$ beamlets. We define $D$ as the lattice constant of the grid and $r_0$ as the initial waist radius of each beamlet. We initialize our input beam to be a coherent and monochromatic ($\lambda$ = 800 nm) continuous wave (CW) beam with a uniform linear polarization throughout, which, without loss of generality, allows us to make use of scalar propagation equations. We assume the nonlinear medium to have a third-order Kerr response with a value of $n_2$ to be the same as fused silica.

For a scalar CW monochromatic field propagating along the $z$-axis, we express the electric field as $E(\mathbf{r},t) = A(x,y,z)e^{{{\rm i}(kz-\omega_0 t)}} +\text{c.c.} $ with $A(x,y;z)$ being the transverse spatial distribution at position $z$.
The non-paraxial propagation of a laser beam inside a Kerr medium is governed by the following equation for field amplitude $A(x,y,z)$ \cite{Feit:88}
\begin{equation}
	\label{np_eq}
	\partial_z^2 A + 2ik \partial_z A  + \nabla_{\bot}^2A + 4\varepsilon_0 c k^2n_2\left|A\right|^2A = 0,
\end{equation}
where $k = \frac{2\pi n_0}{\lambda}$ is the amplitude of the wave vector or the wave number, and $ \nabla_{\bot}^2 = \partial_x^2 + \partial_y^2$ is the transverse Laplacian.
We write the initial field amplitude of a $N\times N$ grid beam at $z=0$ as
\begin{equation}
	A(x,y;z=0) = A_N \Sigma_{n=1}^{N^2}e^{{-\frac{(x-x_n)^2+(y-y_n)^2}{r_0^2}}},
\end{equation}
where $(x_n,y_n)$ is the location of the center of the $n$-th beamlet in the grid and $A_N$ is the amplitude determined
by the input power $P_{\rm input} = 2n_0\varepsilon_0 c\int{\rm d}x{\rm d}y  \left|{A(x,y,0)}\right|^2 $.
Under the appropriate approximations mentioned in \cite{Feit:88}, the solution to Eq. (\ref{np_eq}) over an axial increment  $\Delta z$ becomes
\begin{align}
	\label{sim_eq}
	A(x,y,\Delta z) = &\exp \left\{ \frac{i\Delta z }{2}\left[\frac{ \nabla_{\bot}^2}{\left({ \nabla_{\bot}^2 + k^2}\right)^{1/2} + k}\right]
	\right\}\exp\left[{i k\Delta z  \,(2n_2\varepsilon_0 c)\left|{A(x,y,0)}\right|^2}\right] \notag\\
	&\times\exp \left\{ \frac{i\Delta z }{2}\left[\frac{ \nabla_{\bot}^2}{\left({ \nabla_{\bot}^2 + k^2}\right)^{1/2} + k}\right]
	\right\}
	A(x,y,0) + \mathcal{O}(\Delta z)^3.
\end{align}

We use the split-step-Fourier-method \cite{agrawal2013nonlinear} to solve Eq. (\ref{sim_eq}) assuming the material parameters for fused silica as $n_0 = 1.45$ and $n_2 = 2.9\times 10^{-16}{ \rm cm}^2/{\rm W}$, and the wavelength $\lambda = 800\text{ nm}$. From Equation \ref{P0}, we determine the associated critical power to be 2.3 MW. We then set $r_0$ such that the Rayleigh range $z_{\rm R}$ equals to $114\lambda$ for each beamlet, and perform the simulations in a $1024\times1024$ (or $2048\times2048$ for larger $N$) mesh with a mesh resolution set to  $\Delta x = \Delta y = 0.2 \lambda$ and $\Delta z = 0.25\lambda$. 

\section{Effect of inter-beamlet interaction in a grid structure}
Inter-beamlet interaction plays a central role in determining whether a high-intensity grid beam can propagate stably over long distances in a Kerr medium. Each beamlet produces a local Kerr-induced refractive-index perturbation, which modifies the propagation of its neighboring beamlets and enables lateral energy exchange across the array.

To maximize the power that can be transmitted without collapse over a prescribed distance, we should seek an optimal lattice spacing $D_{\rm optimal}(N, r_0)$ for a $N \times N$ grid beam such that the beam can propagate through a certain distance without collapsing. We define the maximum power that can be stably transmitted (i.e., without collapse) over a distance $z_{\rm max}$ as
\begin{equation}
	\label{beta}
	P_{\rm max}\left(N,D,r_0,{z_{\rm max}}\right) = \beta\left(N,D,r_0,{z_{\rm max}}\right)N^2P_{\rm single}\left({z_{\rm max}}\right),
\end{equation}
where $D$ is the grid lattice spacing and $P_{\rm single}\left(z_{\rm max}\right)$ is the threshold power required for a single \textit{isolated} Gaussian beam to self-focus at $z_{\rm max}$. The dimensionless factor $\beta$ quantifies the net effect of inter-beamlet interaction on the collapse threshold. In the strongly overlapping limit $D\to 0$, the $N^2$ beamlets become indistinguishable and the grid propagates effectively as a single beam, implying $\beta\to\frac{1}{N^2}$. In the opposite limit $D\to \infty$, the beamlets propagate independently and $\,\beta \to 1$.
In our simulations, $\beta$ increases with the grid size $N$ (for $N>2$) and decreases with the beamlet radius $r_0$.

In many practical scenarios the required propagation distance is fixed, so we evaluate collapse behavior over a long but finite distance. As a representative case, we set $z_{\rm max} = 4z_{\rm R}$ and analyze a $4\times4$ grid. For a single Gaussian beam to self-focus by $z_{\rm max} = 4z_{\rm R}$, the power must exceed $P_{\rm single}(z_{\rm max} = 4z_{\rm R}) = 1.041 P_0$ where $P_0$ is the critical power for self-focusing at a finite distance. Equation \ref{beta} then yields $P_{\rm max}(N=4,D,r_0=5\lambda, z_{\rm max} = 4z_{\rm R}) = 4^2 \beta P_{\rm single}(z_{\rm max} = 4z_{\rm R})$ so that $\beta > 1$ directly corresponds to a threshold exceeding the isolated scaling.

\begin{figure}[htbp]
	\centering\includegraphics[width=12cm]{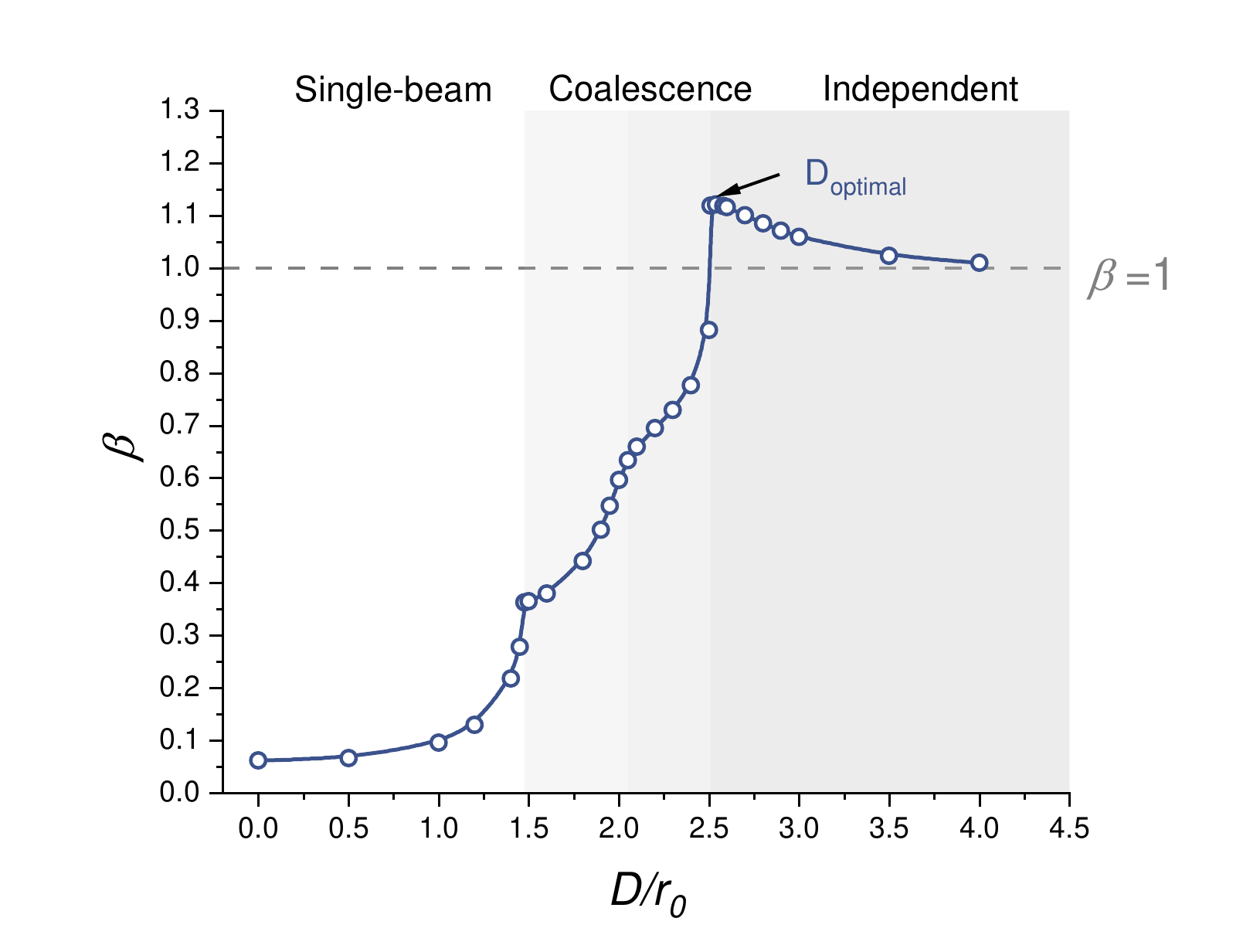}
	\caption{The relation between the power enhancement factor $\beta$ and the lattice constant $D$ for a $4\times4$ grid beam with the maximum propagation distance $z_{\rm max} = 4z_{\rm R}$ and a beamlet radius of $r_0 = 5\lambda$. The enhancement of the threshold power required for self-focusing becomes most pronounced at $D = 2.54r_0$. Note that the four segments of the curve correspond to the four coalescence stages of a $4\times4$ grid beam. The blue dashed line ($\beta = 1$) indicates the asymptotic case where each beamlet propagates independently as the interaction among beamlets is negligible due to a large lattice constant.}
	\label{4g}
\end{figure}

Figure \ref{4g} shows the variation of $\beta(D)$ for the $4\times 4$ grid. The asymptotic behavior at very small and very large $D$ matches the limiting cases discussed above: when $D \rightarrow 0$, the beamlets act as a single beam and collapse once the total power exceeds $P_0$: when $D$ is sufficiently large, the interaction becomes negligible and each beamlet propagates essentially independently. A key feature is a sharp increase in $\beta$ within a narrow spacing interval: as $D$ increases from 2.50$r_0$ to $2.51 r_0$ increases abruptly by 27\% and reaches its maximum value $\beta_{\rm max} = 1.121$. This maximum occurs at $D_{\rm optimal} = 2.54 r_0$ where the power-per-unit-cell ${\beta}/{D^2}$ is also maximized. We observe analogous ``jump-over-unity” behavior in other grid configurations examined in this study, consistent with the physical picture that nonlinear coupling among neighboring beamlets promotes power redistribution (``energy spreading”) and thereby raises the collapse threshold.

\begin{figure}[htbp]
	\centering\includegraphics[width=\textwidth]{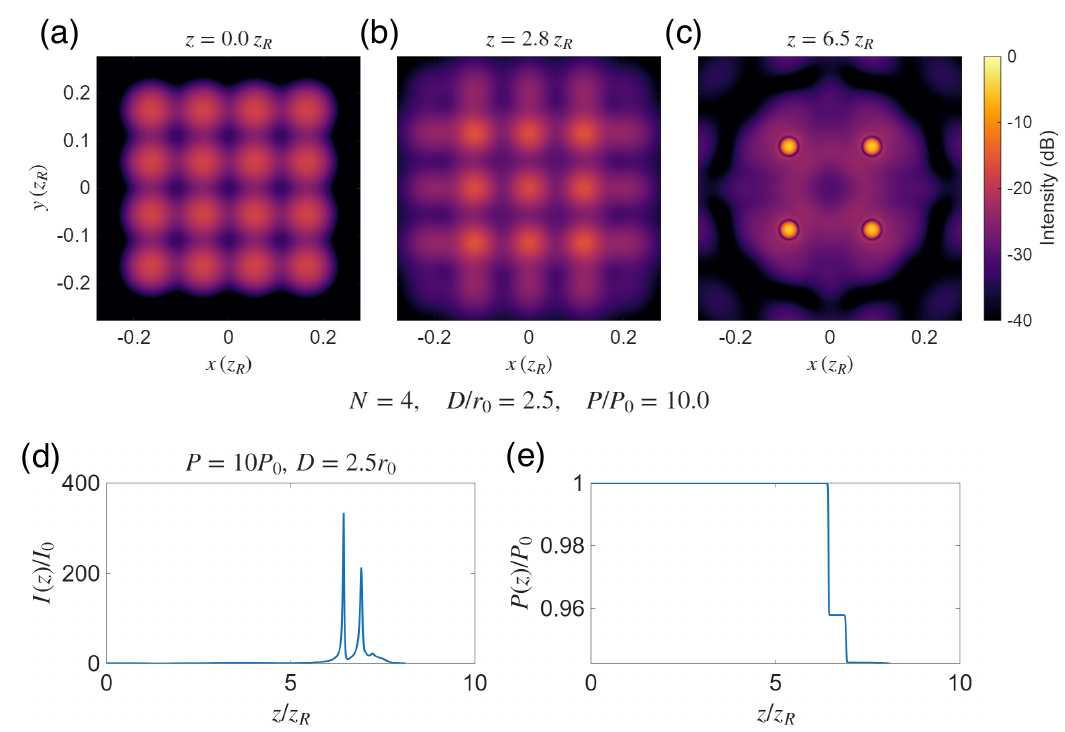}
	\caption{(a) -- (c) Demonstration of a coalescing process of $4\times 4$ grid beam into a (b) $3\times 3$ grid at $z = 2.8z_{\rm R}$ and then into a $2\times 2$ grid at $z = 6.5z_{\rm R}$. (d) Peak intensity and (e) transmitted power over the propagation. Here $P_{\rm input} = 10.0P_0, \text{ }D = 2.50r_0$.}
	\label{coa}
\end{figure}

Beyond increasing the collapse threshold, inter-beamlet interaction produces a distinctive propagation dynamics: an $N\times N$ grid can recursively coalesce into an $(N - 1)\times (N - 1)$ grid, and ultimately into a single beam. For an $N\times N$ grid beam above the collapse threshold, the evolution can be divided into $N-1$ stages. At the $m$-th $(m<N)$ stage, the beam undergoes self-focusing and then continues to propagate as an $(N-m)\times (N-m)$ grid. This multi-stage coalescence reflects the locality of the coupling: each beamlet is influenced predominantly by its nearest neighbors. For the $4\times4$ grid beam, Figure \ref{coa}(a)-(c) show the first two coalescence stages where the incident beam collapses into a $3\times3$ grid and then a $2\times2$ grid during its propagation. The abrupt threshold-like increase in non-collapse power as $D$ crosses a narrow window around $D_{\rm optimal}$ is therefore a signature of a qualitative change in the nonlinear coupling regime (from effectively “single-beam” behavior to a strongly coupled coalescence regime and finally to the independent-beamlet regime at large $D$).

Figures \ref{coa}(d)--(e) show the peak intensity and transmitted energy versus propagation distance for the optimal-spacing configuration. Here, $I_0$ notes the peak intensity at the input plane. We observe three-stage coalescence for the $4\times4$ grid, consistent with the focusing–defocusing cycles reported in \cite{PhysRevLett.76.4356,Feit:88}. The abrupt drops in transmitted energy at self-focusing events arise from evanescent-wave generation near the focus in the non-paraxial model, while the gradual energy decrease is attributable to the hard-absorbing boundary conditions used in the numerical simulation. 

\section{Optimal spacing for 2D square grid and beamlet radius}
\subsection{Dependence on grid size $N$}

\begin{figure}[htbp]
	\centering\includegraphics[width=\textwidth]{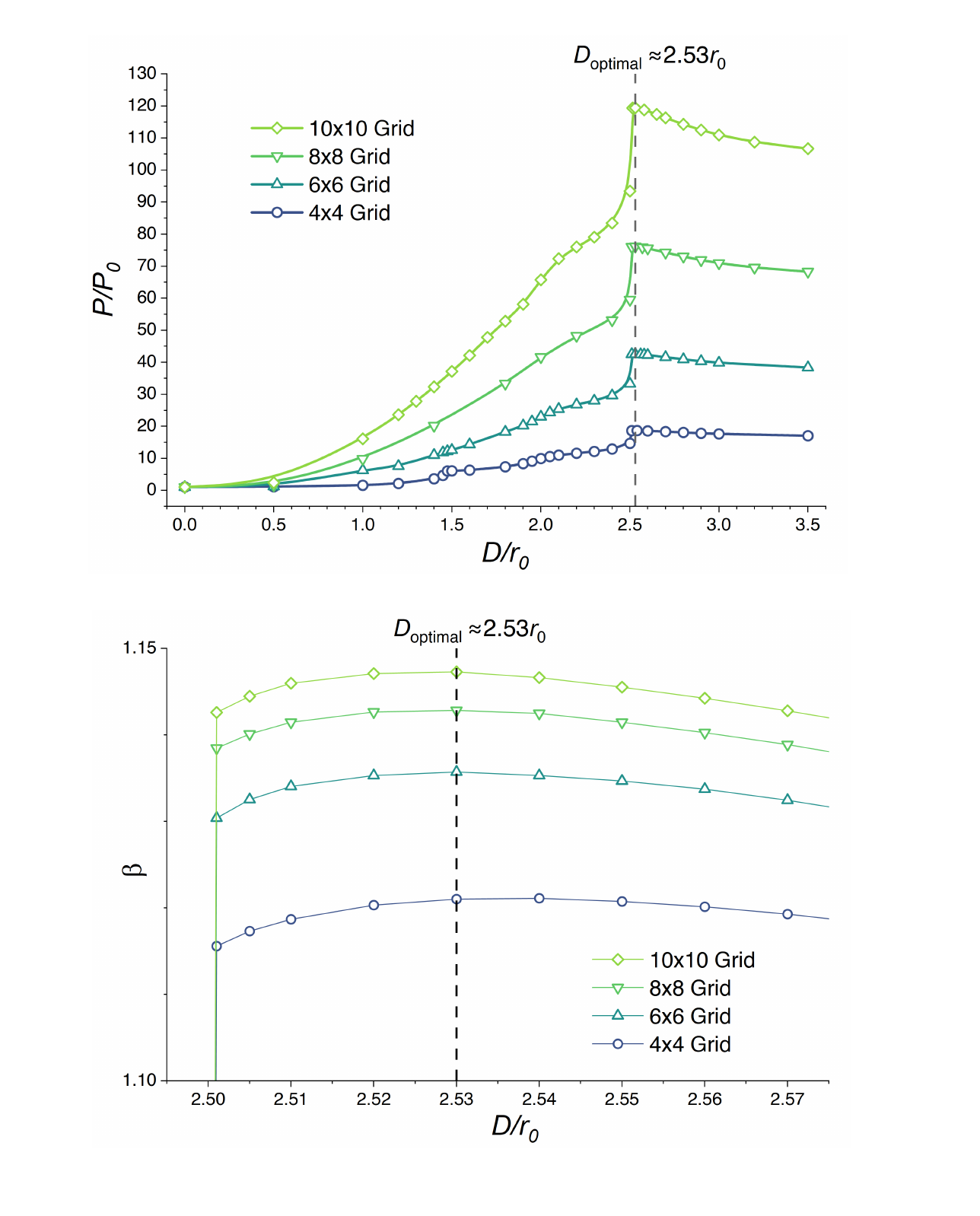}
	\caption{(a) The relation between the threshold power, $P_{\rm max}$, and the lattice spacing, $D$, for grid beams with an even dimension. Notice the significant increase of $P_{\rm max}$ due to the inter-beamlet interaction at the optimal lattice spacing, $D_{\text{optimal}}$. (b) The power enhancement factor $\beta$ near the optimal lattice spacing for even-$N$ grids. The power enhancement factor $\beta = P_{\rm max}/(N^2P_{\rm single})$ uses the fully isolated beamlets as a benchmark to describe how much the inter-beamlet effect can increase the maximum power that can be stably transmitted over $z_{\max}$. The total propagation distance is kept fixed at $z_{\rm max} = 4z_{\rm R}$. For even-$N$ grids, the maximum power enhancement occurs close to $D \approx 2.53r_0$. At $D_{\text{optimal}}$, the collapse of each individual beamlet is delayed because the self-focusing is maximally compensated by the net effect of diffraction and the nonlinear refractive index change induced by surrounding beamlets. }
	\label{odd}
\end{figure}
	\label{even}

We now extend the representative $4\times4$ case to square grids of arbitrary integer dimension $N$. Unless otherwise stated, we assume $r_0 = 5\lambda$ and $z_{\rm max} = 4z_{\rm R}$. For even-$N$ grids, Figure \ref{odd}(a) shows that the maximum transmissible power exhibits a pronounced jump near a common optimal spacing, and Figure \ref{odd}(b) shows that the corresponding enhancement factor $\beta$ is maximized close to $D \approx 2.53r_0$. The extracted optimal separations $D_{\rm optimal}$ and the corresponding peak enhancement factors $\beta_{\rm max}$ are summarized in Table S1 in Supplement 1.

Across all simulated sizes, $D_{\rm optimal}$ follows a simple parity-dependent empirical formula for $z_{\rm max} = 4z_{\rm R}$:
\begin{equation}
        \label{Dopt}
	D_{\rm optimal}(N, r_0, z_{\rm max} = 4z_{\rm R}) \approx 
	\begin{cases}
		2.53r_0, \quad {\text{if}}\; N {\text{ is even}}\\
		2.53 \left(\frac{N-1}{N-2}\right)r_0,\quad {\text{if}}\; N {\text{ is odd}}
	\end{cases}
\end{equation}

The fitted values from Eq.(\ref{Dopt}) agree closely with the simulation results in Table S1 in Supplement 1.
We also find that $\beta_{\max}$ increases monotonically with $N$ and approaches $\sim1.15$ for large grids, indicating that inter-beamlet interaction at the optimal spacing provides a clear enhancement of the non-collapse power relative to the isolated limit.
In addition, odd-$N$ grids require a larger $D_{\rm optimal}$ than even-$N$ grids at fixed $r_0$ and $z_{\rm max}$, while achieving comparable total transmissible power. See Fig. S2 in Supplement 1 for comparison. This odd–even difference weakens with increasing $N$, consistent with Eq. (\ref{Dopt}), where the parity-dependent prefactor approaches the even-$N$ limit for large grids.

\begin{table}[htbp]
	\caption{Optimal lattice constant $D_\text{optimal}$ in terms of the beamlet radius $r_0$ and the maximum power enhancement factor $\beta_{\rm max}$ for various $r_0$ and $N$ at $z_{\rm max} = 4z_{\rm R}$. $D_\text{optimal}/r_0$ remains invariant for different beamlet radii, showing that the optimal configuration depends only on the relative size between the beamlet spacing and its radius.}
    \label{vr}
    \centering
    \resizebox{0.85\textwidth}{!}{%
        \begin{tabular}{c|c|cc|cc|cc|cc}
            \hline
            & & \multicolumn{2}{c|}{$N=2$} & \multicolumn{2}{c|}{$N=3$} & \multicolumn{2}{c|}{$N=4$} & \multicolumn{2}{c}{$N=5$} \\
            \hline
            $r_0/\lambda$ & $P_{\rm single}/P_0$ 
              & $\beta_{\rm max}$ & $D_{\rm opt}/r_0$
              & $\beta_{\rm max}$ & $D_{\rm opt}/r_0$
              & $\beta_{\rm max}$ & $D_{\rm opt}/r_0$
              & $\beta_{\rm max}$ & $D_{\rm opt}/r_0$ \\
            \hline
            3  & 1.047 & 1.121 & 2.45 & 1.107 & 5.09 & 1.119 & 2.53 & 1.129 & 3.37 \\
            5  & 1.041 & 1.108 & 2.52 & 1.109 & 5.08 & 1.121 & 2.54 & 1.129 & 3.37 \\
            7  & 1.039 & 1.123 & 2.45 & 1.110 & 5.07 & 1.121 & 2.54 & 1.130 & 3.38 \\
            10 & 1.038 & 1.130 & 2.43 & 1.110 & 5.08 & 1.122 & 2.53 & 1.130 & 3.39 \\
            \hline
        \end{tabular}
    } \\[1em] 
    \resizebox{0.85\textwidth}{!}{%
        \begin{tabular}{c|c|cc|cc|cc|cc}
            \hline
            & & \multicolumn{2}{c|}{$N=6$} & \multicolumn{2}{c|}{$N=7$} & \multicolumn{2}{c|}{$N=8$} & \multicolumn{2}{c}{$N=9$} \\
            \hline
            $r_0/\lambda$ & $P_{\rm single}/P_0$ 
              & $\beta_{\rm max}$ & $D_{\rm opt}/r_0$
              & $\beta_{\rm max}$ & $D_{\rm opt}/r_0$
              & $\beta_{\rm max}$ & $D_{\rm opt}/r_0$
              & $\beta_{\rm max}$ & $D_{\rm opt}/r_0$ \\
            \hline
            3  & 1.047 & 1.134 & 2.53 & 1.138 & 3.03 & 1.141 & 2.53 & 1.144 & 2.89 \\
            5  & 1.041 & 1.135 & 2.53 & 1.139 & 3.03 & 1.142 & 2.53 & 1.145 & 2.89 \\
            7  & 1.039 & 1.136 & 2.53 & 1.140 & 3.04 & 1.143 & 2.53 & 1.146 & 2.89 \\
            10 & 1.038 & 1.137 & 2.53 & 1.141 & 3.03 & 1.144 & 2.53 & 1.146 & 2.87 \\
            \hline
        \end{tabular}
    }
\end{table}

\subsection{Scaling with beamlet radius $r_0$}

The results above suggest a simple scaling behavior with respect to the beamlet radius $r_0$. To test this idea, we vary $r_0$ while keeping the normalized propagation distance $z_{\max}/z_R$ fixed. The results summarized in Table~1 show that, for even-$N$ grids, the optimal normalized spacing $D_{\mathrm{optimal}}/r_0$ remains essentially invariant for different $r_0$, with similar values of $\beta_{\max}$. For odd-$N$ grids, both $D_{\mathrm{optimal}}/r_0$ and $\beta_{\max}$ are likewise nearly independent of $r_0$.

Further, Eq. (\ref{Dopt}) shows that the optimal lattice spacing $D_{\text{optimal}}$ does not have $N$ dependence for even $N$ $(N=2,4,6\dots)$. This means that the optimal dimensionless normalized lattice spacing, $D_{\text{optimal}}/r_0$, should only depend on the self-focusing distance $z_{\rm max}$ for all even-$N$ grids. Since the beam width scaling only affects the Rayleigh range, this suggests that if we keep $\frac{z_{\rm max}}{z_{\rm R}}$ fixed while changing $r_0$, the propagation behavior of grid beams inside the same nonlinear medium should also be invariant up to the scaling factor. As such, the critical power depends on the number of beamlets, and not on the radius of the individual beamlet \cite{boyd2020nonlinear,cmp/1103922134,Fibich:00}. This scaling invariance substantiates the common understanding that self-focusing behavior depends on the power of the beam instead of its intensity. As shown in Table \ref{vr}, we vary the input beam radius $r_0$ and have verified that for different $r_0$, all even-$N$ grids have the same ${D_{\rm optimal}}/{r_0}$ and similar maximum power enhancement factors $\beta_{\rm max}$. For odd-$N$ grids, we find that both ${D_{\rm optimal}}/{r_0}$ and $\beta_{\rm max}$ are independent of $r_0$. 

\subsection{Dependence on propagation distance $z_{\max}$}

\begin{figure}[htbp]
	\centering\includegraphics[width=0.9\textwidth]{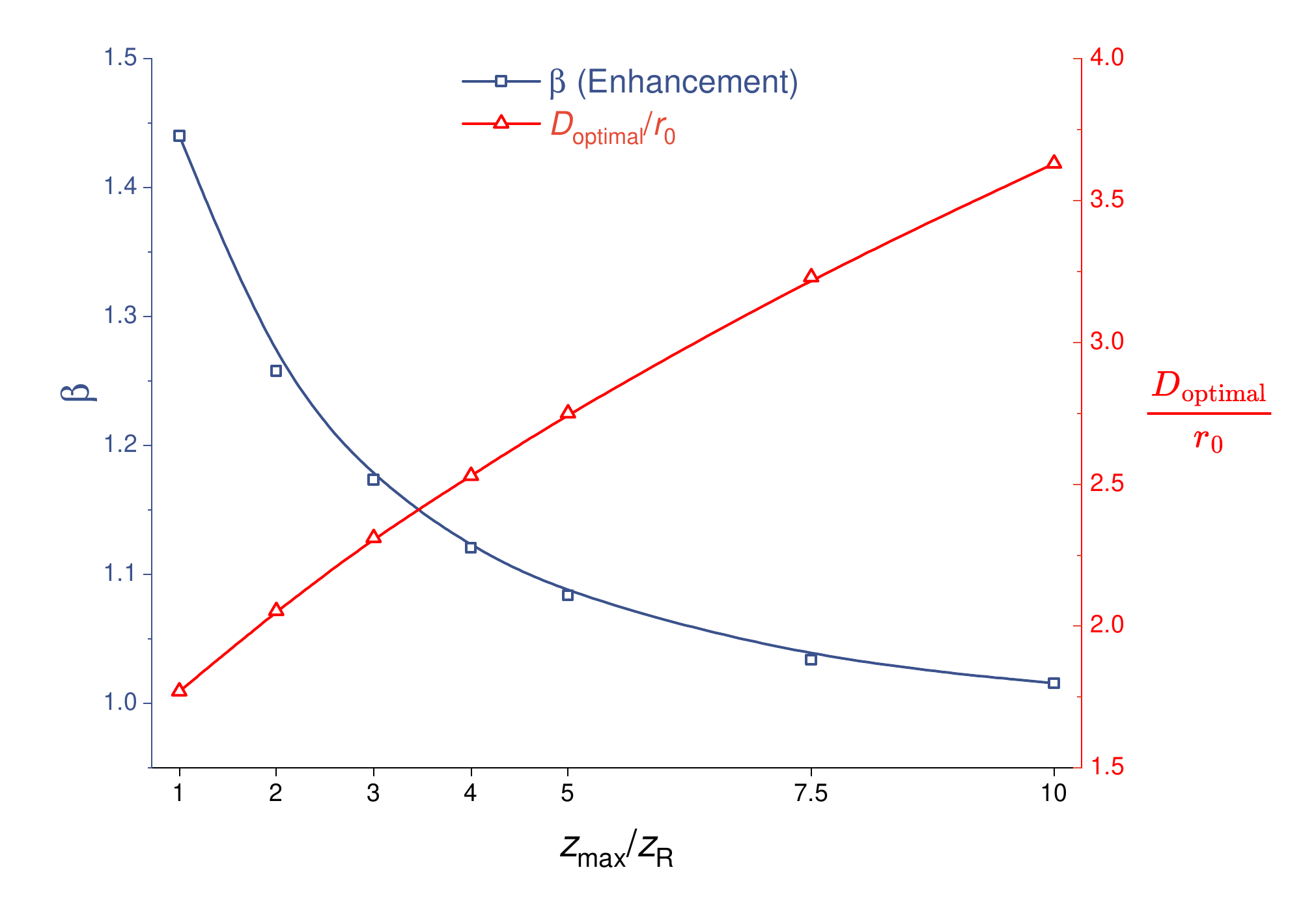}
	\caption{Optimal lattice constant $D_\text{optimal}$ normalized to the beamlets radius $r_0$ (drawn in red) and the maximum power enhancement factor $\beta_{\rm max}$ (drawn in blue) of a $4\times 4$ grid for different distances $z_{\rm max}$. At $z_{\rm max} = z_{\rm R}$, a 1.4-fold power enhancement is observed at the optimal lattice spacing. The maximum power enhancement $\beta_{\rm max}$ decreases as the propagation distance grows, whereas the optimal lattice spacing $D_{\rm optimal}/r_0$ increases. The $4\times 4$ grid tends to behave like a matrix of independently propagating beamlet with no inter-beamlet interaction when $z_{\rm max}$ becomes large enough. The concavity of the $\beta_{\rm max}$ curve indicates that the grid structuring manifests a stronger inter-beamlet interference at a shorter propagation distance. }
	\label{zmax_dependence}
\end{figure}

Finally, we examine how the optimum grid parameters depend on the required propagation distance. We have performed numerical self-focusing simulations of a $4\times 4$ grid to determine the optimal lattice spacings and their associated power enhancement for $z_{\rm max}$ up to $10z_{\rm R}$. Figure \ref{zmax_dependence} shows the maximum enhancement factor $\beta_{\rm max}$ and the optimal lattice spacing $D_{\rm optimal}$ for different self-focusing distances $z_{\rm max}$. As expected, the maximum power enhancement factor $\beta_{\max}$ that can be transmitted without collapse decreases and gradually approaches 1 as $z_{\rm max}$ increases. On the other hand, the neighboring beamlets should be pushed further apart as $z_{\rm max}$ grows. At $z_{\rm max} = z_{\rm R}$, the inter-beamlet interaction in a $4\times 4$ grid leads to a 44\% increase in the self-focusing power threshold at the optimal lattice spacing. The concavity of the power enhancement curve in Figure \ref{zmax_dependence} (shown in blue) suggests that the inter-beamlet effect becomes more important in determining the self-focusing behavior at a shorter propagation distance. 

\section{Discussion and conclusion}

In summary, we demonstrate that a 2D periodic beamlet grid can be substantially more robust against catastrophic self-focusing than a single Gaussian beam carrying the same total power. By arranging independent beamlets into a properly spaced square lattice, the nonlinear propagation proceeds through cooperative inter-beamlet interaction that redistributes power transversely and delays localized collapse. As a result, the grid can transmit more power than the fully separated-beamlet limit $N^2P_{\rm single}$. This enhancement is strongly distance-dependent: for a $4\times4$ grid, the total power that can be transmitted becomes 44\% larger for $z_{max} = z_{\rm R}$, and remains 15\% larger for $z_{max} = 4z_{\rm R}$, than a grid with isolated beamlets. These results show that transverse spatial structuring provides a passive and effective way to improve beam robustness against catastrophic collapse over a finite propagation distance.

Furthermore, we identify a simple spatial scaling behavior that explains the nonlinear evolution of these 2D scalar grid beams. For a fixed $z_{\max}/z_R$, the optimal normalized lattice spacing $D_{\mathrm{optimal}}/r_0$ is invariant for even-$N$ grids and effectively invariant with respect to changes in the beamlet radius $r_0$; the maximum enhancement factor $\beta_{\max}$ is likewise only weakly dependent on $r_0$ for both even- and odd-$N$ cases. This behavior supports the power-dominated, rather than intensity-dominated, character of the collapse dynamics in the present system. It is also practically valuable, because the computational volume scales steeply as $z_R r_0^2 \propto r_0^4$. Therefore, one can perform simulations at smaller beamlet size and reliably infer the optimum grid behavior at larger scales, which significantly reduces the numerical cost of parameter optimization.

Ultimately, the proposed beam-structuring strategy offers a practical and passive route to enhanced stability in nonlinear propagation. By mitigating catastrophic self-focusing through structured beam engineering, this approach may be useful for high-power optical beam delivery and directed-energy applications. Accordingly, our numerical model can be further extended to include higher-order nonlinearities and full vectorial propagation.

\begin{backmatter}
\bmsection{Funding}
Content in the funding section will be generated entirely from details submitted to Prism. Authors may add placeholder text in this section to assess length, but any text added to this section will be replaced during production and will display official funder names along with any grant numbers provided. If additional details about a funder are required, they may be added to the Acknowledgment, even if this duplicates some information in the funding section. For preprint submissions, please include funder names and grant numbers in the manuscript.

\bmsection{Disclosures}
The authors declare no conflicts of interest.

\bmsection{Data availability} Data underlying the results presented in this paper are not publicly available at this time but may be obtained from the authors upon reasonable request.

\bmsection{Supplemental document}
See Supplement 1 for supporting content.

\end{backmatter}

\bibliography{grid_beam}
\end{document}